\documentclass[12pt]{article}

\usepackage{graphicx,amssymb,lineno}

\begin{document}

\topmargin -2pt

\headheight 0pt

\begin{center}

{\Large \bf  How does an external electrical field affect adsorption patterns of thiol and thiolate on the gold substrate ?}\\

\vspace{10mm}

{\large Jian-Ge Zhou$^{a,b}$, Quinton L. Williams$^{b}$}\\
{\it $^{a}$Center for Molecular Structure and Interactions, Jackson 
State University, Jackson, MS 39217, USA}\\
{\it $^{b}$Department of Physics, Atmospheric Sciences, and Geoscience, Jackson 
State University, Jackson, MS 39217, USA}\\

\vspace{10mm}

{\bf ABSTRACT} \\
\end{center}

\vspace{2mm}

\noindent
The responsive behavior of methanethiol and methylthiolate molecules on the Au(111) surface with an applied electrical potential
is studied, and it is shown 
how the sulfur adsorption site, the S-H bond orientation and the interacting energy change with
an external electric field strength. 
The electron charge density corresponding to an electric
field minus that obtained in zero field, with zero-field optimal geometry,
is calculated to explain the responsive behavior. The 
interacting energy for the intact methanethiol adsorption is larger than
that for the dissociative one, showing that an external electric
field can not make the hydrogen dissociate from the sulfur. 
\\\\


\newpage
%%%%%%%%%%%%%%%%%%%%%%%%%%%%%%%%%%%%%%%%%%%%%%%%%%%%%
%%%%%%%%%%%%%%%%%%%%
\section{Introduction}

Materials and devices that change properties and functions in response to external
stimuli are the focus of research in fields of physics, chemistry, biology, material
science and engineering \cite{wdi}-\cite{gk}. Physical effects such as external fields are advantageous 
in the process of controlling surface adsorption and growth.
As we know, the surface morphology can be easily affected by an external electric field. 
The potential-induced surface morphological changes are observed in metal/electrolyte
interface \cite{wdi}. 
An excess surface charge can induce a reconstruction on a silver surface \cite{fh}.
Adsorbates on the surface are stabilized by the presence of the scanning tunneling microscopy
(STM) tip \cite{ssj}. 
An electric field or surface charging changes the metal bcc(100) surface
configurations \cite{czc}. 
The electric field effects on surface diffusion has been studied by the
field ion microscopy (FIM) technique, and it was found that an electric field can 
inhibit or promote surface self-diffusion on Pt(001) surface \cite{gk}.

On the other hand, alkanethiols form self-assembled monolyers (SAM) on the
Au(111) surface, which has wide applications in molecular electronics \cite{jga}, lubrication \cite{jth},
lithography \cite{lxq}, and bio-chemical surface functionalization \cite{mm}. Its highly ordered
structures and chemical stability make these systems ideal for study with a variety of 
techniques including atomic-force microscopy \cite{ths}, infrared spectroscopy \cite{bw,nkd}, high-resolution
electron-energy-loss spectroscopy \cite{kcm}, grazing X-ray diffraction \cite{fee}, scanning probe
microscopy \cite{asp}, low-energy electron diffraction \cite{nc1},
STM \cite{pt}-\cite{ywu} and others \cite{sf}-\cite{vbg}. Recently, Maksymovych and
coworkers exploited the STM tip to manipulate the formation and decomposition
of the methanethiol dimer on the Au(111) surface \cite{msd}, which shows that an external 
electric field does affect the adsorption pattern of a thiol molecule.
Then, the question of whether an external electrical field 
induces conformal reorientation of thiol molecule on the Au(111)  surface at the low coverage arises.
Because of its importance for a wide variety of surface phenomena (i.e., STM, FIM and electrochemical), understanding
the influence of an external electric field on surface adsorption is essential for
explaining some experimental results. Besides these, the defect on the substrate can catalyze the S-H bond breaking in the process of the 
methanethiol adsorption on the Au(111) surface \cite{rlm,zh}, however, 
it is unclear if an external electric field can trigger such a dissociation.
Heretofore, the mechanism of the responsive behavior and the dissociation 
of thiol molecule on the Au(111) surface
under an external electrical field is still a mystery. 

This prompted us to investigate the interacting behavior of methanethiol 
and methylthiolate molecules 
with the Au(111) surface under an external electrical field 
by the density functional theory.
We will present the interacting energies and geometries for methanethiol and
methylthiolate adsorbates on the Au(111) surface in the presence of an external electric field. 
We show how the sulfur adsorption site, the S-H bond orientation, and the interacting energy 
of the methanethiol and methylthiolate molecules with the Au(111) substrate are affected
by an external electric field applied to the surface. 
We have calculated the z-direction electron charge density difference
between the charge density obtained with an electric field and that without a field
at the zero-field optimized geometry to interpret these responsive 
behaviors. To see if an external electric field can trigger the dissociation of the
S-H bond in the methanethiol adsorbed on the Au(111) surface, 
we compare the interacting energies between the 
intact adsorption and dissociative one.
We find that the interacting energy for the intact methanethiol adsorption is larger than
that for the dissociative adsorption, which shows that even in the presence of 
an external electric field, the intact adsorption is still stable. 

\section{Theoretial Method and Surface Modeling}

The calculations were done in the slab model by
density functional theory (DFT) \cite{khf}. The electron-ion interaction has been described using the projector
augmented wave (PAW) method.
All calculations have been performed by Perdew-Wang 91 (PW91) generalized gradient approximation.
The wave functions are expanded in a plane wave basis with an energy cutoff of 400 eV. The $k$ points
were obtained from Monkhorst-Pack scheme, and $3\times 3\times 1$  $k$ point mesh was 
for the geometry optimization.
The supercell consisted of 4 layers and each layer with 12 Au atoms.
The Au atoms in the top three atomic layers are allowed to relax, 
while those in the bottom layer are fixed to simulate bulk-like termination \cite{zwh}. 
The vacuum region comprises seven atomic layers, which
exceeds substantially the extension of the methanethiol molecule. To apply an
external electrical field, a planar dipole layer is placed in the middle of the
vacuum region \cite{khf,ns}. 
In the presence of an external electrical field, the eight Au layers slab resulted in charge sloshing. 
We also compared the six layers slab with the four layers slab, and 
found that the differences of the interacting energy are with 5.3$\%$. However, the computing time for the six layers slab is much longer than that for the four
layers slab. In our work, we computed more than 150 configurations, so the best choice for us is the four layers slab.
We calculated the gold lattice constant and found it to agree with the experimental value \cite{ak} 
to 2.1$\%$.

\section{Results and Discussion}

\begin{figure}
\begin{center}
\includegraphics[width=4.5in]{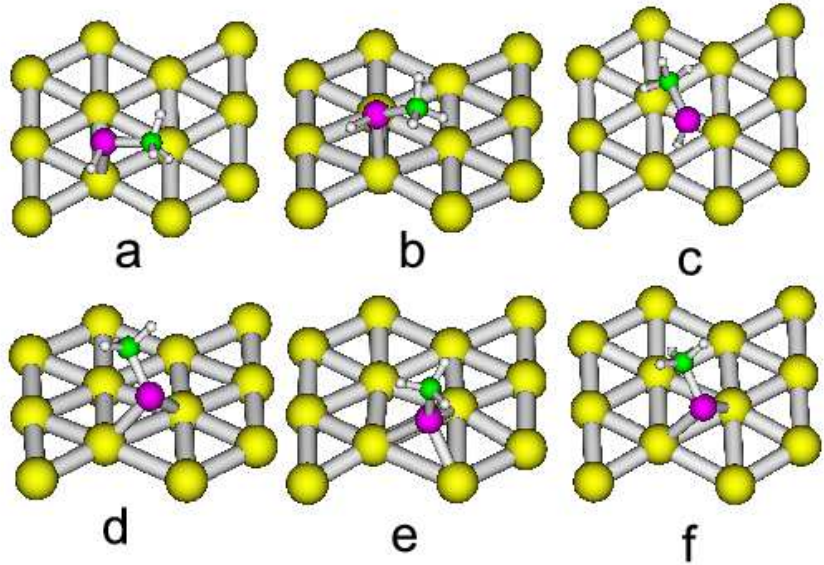}
\end{center}
\caption{(a) The methanethiol (CH$_3$SH) on the Au(111) surface without an external electric field. 
(b) CH$_3$SH on the surface with a negative external electric field (-1.0 $V/\AA$).
(c) CH$_3$SH on the surface with a positive external electric field (0.5 $V/\AA$). 
(d) Methylthiolate (CH$_3$S) on the Au(111) surface without external electric field. 
(e) CH$_3$S on the surface with a negative external electric field (-1.0 $V/\AA$).
(f) CH$_3$S on the surface with a positive external electric field (1.0 $V/\AA$). } 
\label{structure}
\end{figure}

We begin with the geometries and interacting 
energies of the optimized structures for the methanethiol (CH$_3$SH) on the Au(111) surface  
at the coverage of 0.25 ML (1.00 ML  means 1 sulfur per 3 gold atoms, and 0.25ML stand for 1 methanethiol on a gold surface with 12 gold atoms)
with various external electric field strengths \cite{ns}, as displayed 
in Table \ref{geom} (at each value of the external electric field, 15 different structures have been optimized, the most
stable structure is listed on table \ref{geom}).
%%%%%%%%%%%%%%%%%%%%%%%%%%%%%%%%%%%%%%%%%%%%%%%%%%%%%%%%%%%%%%%%%%%%%%%%%%%%%%%%%%%%%%%%
\begin{table}
\caption{The geometries and interacting energies for the stable methanethiol configurations 
on the Au(111) surface (0.25 ML) at various external electric field strengths. The entries E$_{ext}$, S site,
$\theta$, tilt, $d_{S-Au}$ ($\AA$) and $E_{int}$ (eV) refer to an external 
electric field (V/$\AA$) perpendicular to the Au(111) surface, the S atom adsorption
site, the angle between the S-C bond direction and the normal to the Au(111) surface, the region
of the S-C bond tilted,  
the shortest S-Au bond length and the interacting energy, respectively.}
\begin{center}
\begin{tabular}{cccccc}  
\hline
\hline
E$_{ext}$ &S site &$\theta$&tilt& $d_{S-Au}$ & $E_{int}$\\
\hline
0&top-fcc&73.0 &fcc&2.73&0.66 \\
-0.5&top-fcc&72.5&fcc&2.61&0.96\\
-1.0&top-fcc&65.0&fcc&2.57&1.41\\
-1.5&top-fcc&54.1&fcc&2.55&1.80\\
0.5&bri&97.4 &hcp&3.95&0.50\\
\hline
\end{tabular}
\end{center}
\label{geom}
\end{table}
%%%%%%%%%%%%%%%%%%%%%%%%%%%%%%%%%%%%%%%%%%%%%%%%%%%%%%%%%%%%%%%%%%%%%%%%%%%%%%%%%%%%%%%%
The interacting energy is defined as $E_{int}$ = $E_{CH3SH}$ + $E_{Au(111)+field}$ - $E_{CH3SH+Au(111)+field}$.
The symbol top-fcc (or top-hcp) in Table \ref{geom} represents that the S atom 
is at the atop site of the gold atom, but leaned toward the fcc (or hcp) hollow center.
Some stable configurations on the Au(111) surface in the presence of an
external electric field are illustrated in Fig. \ref{structure}, where only the methanethiol (or
methylthiolate) adsorbate and the top layer of the Au(111) surface are displayed.

Table \ref{geom} shows that when the strength of an applied negative electric field increases, 
the interacting energy E$_{int}$ rises, the sulfur adsorption site shows
little variation (on the atop site of the gold atom, but leaned to fcc center), 
the angle between the S-C bond direction and the normal to the Au(111) surface decreases, and
the bond length between S atom and substrate $d_{S-Au}$ becomes shorter. Thus, when the 
strength of an applied
negative electric field increases, the interaction between the methanethiol adsorbate and
gold substrate gets stronger and stronger. 
If the negative electric field is in the
range of 0 to -0.5 $V/\AA$, the geometry changes slightly, which is in accord with the
experimental observation \cite{htg,our}. 
When a positive electric field (0.5 $V/\AA$) applied,
the methanethiol molecule starts to desorb from the Au(111) surface.  Fig. \ref{structure}c 
depicts this desorption structure in which the distance between S and Au is 3.95 $\AA$ (longer
than the zero-field S-Au bond length 2.73 $\AA$). Thus in the low coverage, the orientation of the methanethiol
molecule on the Au(111) surface can be tuned by an applied negative electrical field in a certain 
range (-0.5 $V/\AA$ to -1.5 $V/\AA$).

%%%%%%%%%%%%%%%%%%%%%%%%%%%%%%%%%%%%%%%%%%%%%%%%%%%%%%%%%%%%%%%%%%%%%%%%%%%%%%%%%%%%%%%%
 %%%%%%%%%%%%%%%%%%%%%%%%%%%%%%%%%%%%%%%%%%%%%%%%%%%%%%%%%%%%%%%%%%%%%%%%%%%%%%%%%%%%%%%%

To see how an external electric field influences the interaction between the methylthiolate
molecule (CH$_3$S) and the substrate, we calculated the geometries and interacting 
energies for the optimized structures of the methylthiolate on the Au(111) surface  
(0.25 ML) at various external electric field strengths, as shown in Table \ref{geom1}. 
%%%%%%%%%%%%%%%%%%%%%%%%%%%%%%%%%%%%%%%%%%%%%%%%%%%%%%%%%%%%%%%%%%%%%%%%%%%%%%%%%%%%%%%%
\begin{table}
\caption{The geometries and interacting energies for the stable methylthiolate configurations 
on the Au(111) surface (0.25 ML) at various external electric field strengths.}
\begin{center}
\begin{tabular}{cccccc}  
\hline
\hline
E$_{ext}$ &S site &$\theta$&tilt& $d_{S-Au}$ & $E_{int}$\\
\hline
0&fcc-bri&55.6 &hcp&2.45&2.31 \\
-0.5&fcc-bri&55.3 &hcp&2.45&2.48\\
-1.0&fcc&1.1 &hcp&2.46&2.83\\
-1.5&fcc&1.0 &hcp&2.45&3.07\\
0.5&fcc-bri&57.1 &hcp&2.48&2.22\\
1.0&fcc-bri &61.9 &hcp&2.52&2.22\\
1.5&fcc-bri &58.0 &hcp&2.50&2.28\\
\hline
\end{tabular}
\end{center}
\label{geom1}
\end{table}
%%%%%%%%%%%%%%%%%%%%%%%%%%%%%%%%%%%%%%%%%%%%%%%%%%%%%%%%%%%%%%%%%%%%%%%%%%%%%%%%%%%%%%%%
Table \ref{geom1} displays that if the strength of an applied negative electric field becomes 
stronger, the interacting energy increases, the sulfur adsorption site is sliding from fcc-bri to fcc, 
the angle $\theta$ decreases, but
the bond length $d_{S-Au}$ shows little variation. 
When the electric field goes to -1.0 $V/\AA$, the
previous tilted methylthiolate molecule begins to stand up, i.e., the angle $\theta$ jumps from $55^\circ$
to $1^\circ$. When the strength of a
negative electric field increases, the interaction between the methanethiol adsorbate and
gold substrate gets stronger, but the bond length $d_{S-Au}$ remains unchanged.  
When applying a positive electric field, if the field strength increases, 
the interacting energy first decreases then increases. 
Unlike the methanethiol case, the S-Au distance in the methylthiolate adsorbate
with a positive potential is near to that with zero field.  Fig. \ref{structure}f reveals that 
the methylthiolate adsorption structure on the Au(111) surface with a positive external
electrical field looks like that without an external electrical field (Fig. \ref{structure}d).  
The calculation shows that we cannot adjust the orientation of the methylthiolate on the
Au(111) surface continuously. In the range of -0.5 to 1.5 $V/\AA$, the angle of the S-C bond is around
$55^\circ$, but within -1.0 to -1.5 $V/\AA$, the methylthiolate is nearly vertical to
the surface. When the negative electric field is in the
range of 0 to -0.5 $V/\AA$, even when the interacting
energy varies, the orientation of the methylthiolate molecule
almost does not change, which is consistent with the
experimental results \cite{ebk}.

%%%%%%%%%%%%%%%%%%%%%%%%%%%%%%%%%%%%%%%%%%%%%%%%%%%%%%%%%%%%%%%%%%%%%%%%%%%%%%%%%%%%%%%%
\begin{figure}
\begin{center}
\includegraphics[width=6.0in]{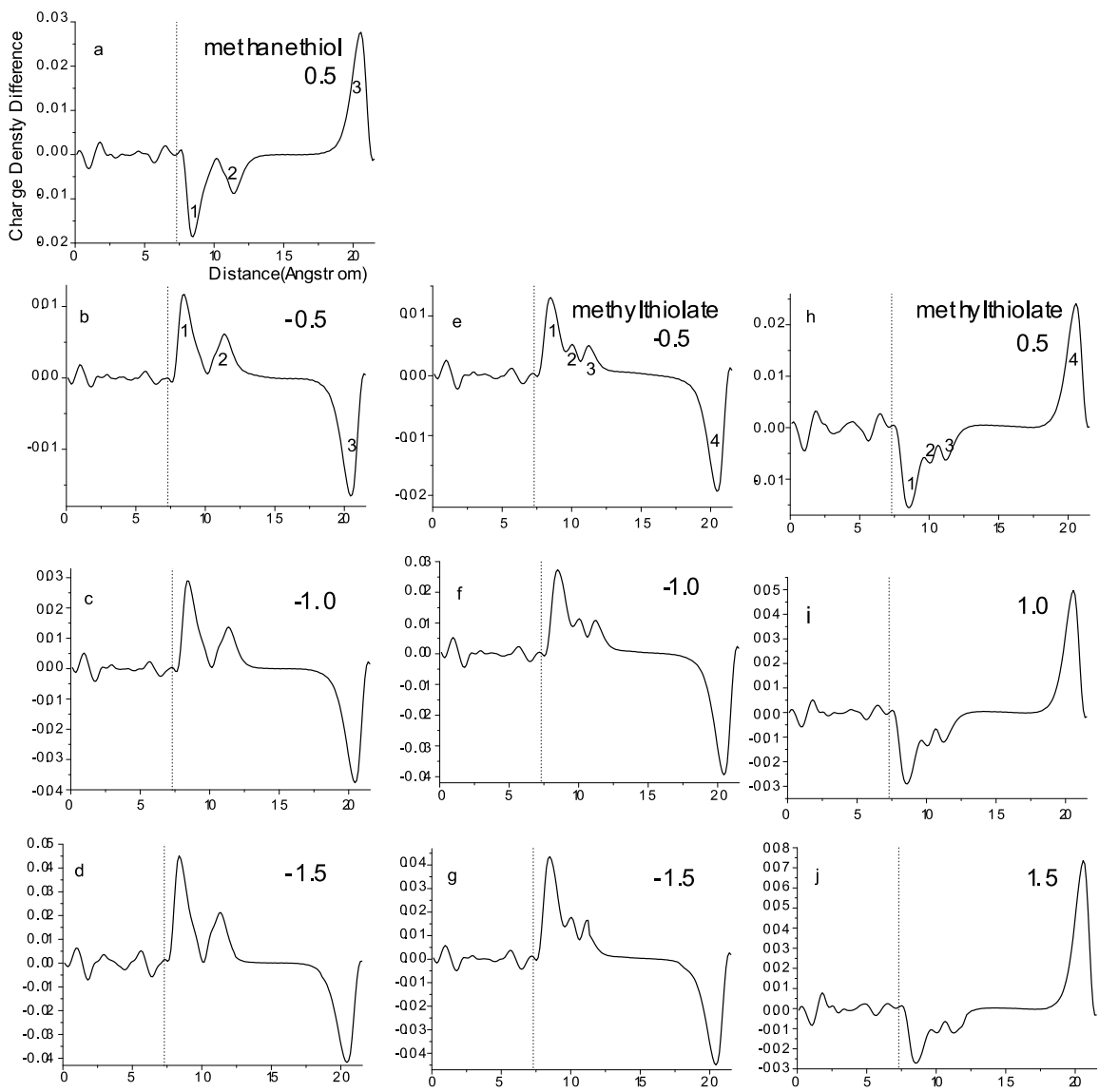}
\end{center}
\caption{The electron charge density difference along the surface normal defined as the 
the charge density corresponding to an electric
field minus that obtained in zero field, with zero-field optimal geometry. (a)-(d) methanethiol, (e)-(g) methylthiolate
with the negative electric field, and (h)-(j) methylthiolate with the positive electric
field. The vertical dotted line represents the position of the gold top layer.} 
\label{diffchg}
\end{figure} %%%%%%%%%%%%%%%%%%%%%%%%%%%%%%%%%%%%%%%%%%%%%%%%%%%%%%%%%%%%%%%%%%%%%%%%%%%%%%%%%%%%%%%%
Let us calculate the electron charge density  
difference along the surface normal to interpret the responsive behavior. The charge density subtraction is
between the charge density obtained with an electric field and that without an electric field at the 
zero-field optimized geometry.
We have plotted the plane-integrated charge density difference as a function of the 
z-coordinate (Fig. \ref{diffchg}), which shows how the charges rearrange on application of an 
external electric field. 
In the case of the methanethiol adsorption, the positive electric field pulls the electrons back to the gold surface.
Troughs 1 and 2 in Fig. \ref{diffchg}a indicate the removal
of the electrons from the region between the gold surface and sulfur (trough 1) and that
between sulfur and $CH_3$ methyl group (trough 2). The corresponding S-Au bond becomes weaker 
and more electrons have accumulated on the other side of the slab (peak 3 in Fig. \ref{diffchg}a) 
than in cases without an electric field. 
The peaks 1 and 2 in Fig. \ref{diffchg}b display that the negative field
pushes more electrons into the region between S and Au (peak 1) and that around $CH_3$ methyl
group (peak 2). The S-Au bond gets stronger than that without an electric field. In the presence of a negative
electric field, the negatively charged methyl
group tends to move away from the surface. However, table \ref{geom} shows that
the S-C bond length changes slightly in a negative electrical field. Thus, the net effect
is that when the strength of an applied negative electrical field increases (Fig. \ref{diffchg}b - Fig. \ref{diffchg}d), 
the angle between the S-C bond and the surface normal decreases, which 
explains the responsive behavior of the angle $\theta$.
The methylthiolate adsorption is similar to the methanethiol case. 
In an applied negative electrical field, there are
more electrons accumulated around the methyl group (CH$_3$) in the methylthiolate
than in the methanethiol (Fig. \ref{diffchg}b - Fig. \ref{diffchg}d and Fig. \ref{diffchg}e - Fig. \ref{diffchg}g).
When the amount of the electron accumulation exceeds a certain level, the methylthiolate 
becomes nearly vertical to the surface. 
In a positive potential, some electrons flow back to the gold surface (Fig. \ref{diffchg}h - Fig. \ref{diffchg}j); 
the S-Au bond in methylthiolate
gets weaker than that without an electric field. Thus, we have shown how the system responds geometrically 
to the rearrangement of charges in the presence of an applied field.

%%%%%%%%%%%%%%%%%%%%%%%%%%%%%%%%%%%%%%%%%%%%%%%%%%%%%%%%%%%%%%%%%%%%%%%%%%%%%%%%%%%%%%%%
\begin{figure}
\begin{center}
\includegraphics[width=3.2in]{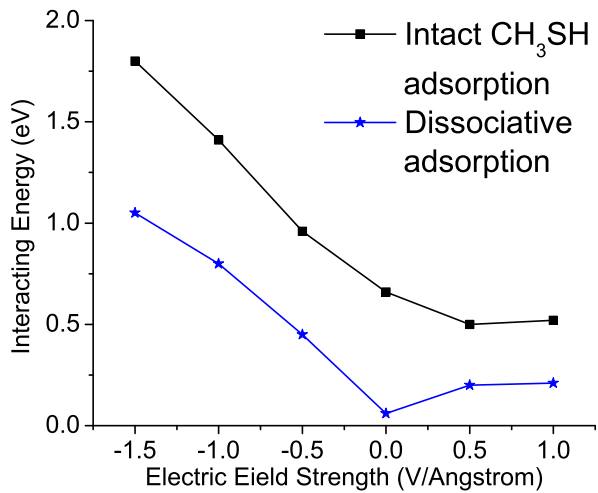}
\end{center}
\caption{The interacting energies for the intact methanethiol adsorption and dissociative adsorption.} 
\label{energydiff}
\end{figure} %%%%%%%%%%%%%%%%%%%%%%%%%%%%%%%%%%%%%%%%%%%%%%%%%%%%%%%%%%%%%%%%%%%%%%%%%%%%%%%%%%%%%%%%
When the methanethiol is adsorbed on the Au(111) surface, the S-H bond remains intact \cite{rlm}. If the
temperature rises, the methanethiol will desorb from the surface. To see if an applied 
electric field can break the S-H bond of the methanethiol adsorbate, we calculated the
interacting energies for the stable structures of the intact (CH$_3$SH) and dissociative adsorption (CH$_3$S + H-Au)
on the Au(111) surface. The interacting energies for the intact 
and dissociative adsorption versus the electric field are plotted in Fig. \ref{energydiff}.
 Fig. \ref{energydiff} reveals that from -1.5 to 0.5 $V/\AA$, the interacting energy
for intact methanethiol adsorption decreases and from 0.5 to 1.0 $V/\AA$, it increases.
In the case of dissociative adsorption,  from -1.5 to 0.0 $V/\AA$, the interacting energy decreases;
however, above 0.0 $V/\AA$, the interacting energy increases. 
Fig. \ref{energydiff} displays that in the whole region, the interacting energy for the intact methanethiol adsorption is larger than
that for the dissociative one, i.e., the intact adsorption
is more stable than the dissociative one. This shows that an external electric
field cannot make the hydrogen dissociate from the sulfur. 

\section{Conclusion}

Based on ab initio calculations, we have shown for the first time how the methanethiol 
and methanethiolate molecules on the Au(111) surface respond to an applied electrical 
potential. 
The sulfur adsorption site, the S-H bond orientation, and the interacting energy 
vary with the strength of the external electric field. 
In the low coverage, the orientation of the methanethiol
molecule on the Au(111) surface can be tuned by the application of a negative electrical field 
through a certain range
and the methanethiol desorbs from the gold substrate with a positive electrical field.
However, the orientation of the methylthiolate on the
Au(111) surface cannot be adjusted continuously. 
The electron charge density (along the surface normal) corresponding to the external
field minus that obtained in zero field, with zero-field optimal geometry, has been calculated 
to interpret these responsive behaviors. The interacting energies between the 
intact and dissociative adsorption with an applied electrical potential have been compared. It has been
found that the interacting energy for the intact methanethiol adsorption is larger than
that for the dissociative adsorption, showing that an external electric
field cannot make the hydrogen dissociate from the sulfur. 

\section*{Acknowledgments}

This work is funded in part by Department of Defense 
through Contract $\#$W912HZ-06-C-0057.

% The Appendices part is started with the command \appendix;
% appendix sections are then done as normal sections
% \appendix

% \section{}
% \label{}


\begin{thebibliography}{00}
\bibitem{our} Upon applying a voltage with respect to a counter electrode, a diffuse layer of ions is formed. 
However, the interaction between the gold electrode and solvent molecules is small, 
the model can be regarded as a first-order approximation \cite{lmt}.
\end{thebibliography}
\end{document}